\documentclass[a4paper,11pt]{article}
\usepackage{pos}
\usepackage{siunitx}
\usepackage{csquotes}
\usepackage{amsmath}
\usepackage{url}
\usepackage{aas_macros}
\usepackage{physics}

\title{TELAMON: Effelsberg Monitoring of AGN Jets with Very-High-Energy Astroparticle Emissions - Polarization properties}
 \ShortTitle{Polarization properties of TeV-blazars in the TELAMON sample}

\author*[a,b]{J.~Heßdörfer}

\author[a]{M.~Kadler}
\author[b,a]{P.~Benke}
\author[b]{L.~Debbrecht}
\author[a]{J.~Eich}
\author[a,b]{F.~ Eppel}
\author[c]{A.~Gokus}
\author[d]{S.~Hämmerich}
\author[a]{D.~Kirchner}
\author[b]{G.F.~Paraschos}
\author[a,b]{F.~Rösch}
\author[a]{W.~Schulga}
\author[e]{J.~Sinapius}
\author[a]{P.~Weber}

\affiliation[a]{Julius-Maximilians-Universität Würzburg, Fakultät für Physik und Astronomie, Institut für Theoretische Physik und Astrophysik, Lehrstuhl für Astronomie, Emil-Fischer-Str. 31, D-97074 Würzburg, Germany}
\affiliation[b]{Max-Planck-Institut für Radioastronomie, Auf dem Hügel 69, D-53121, Bonn, Germany}
\affiliation[c]{Department of Physics \& McDonnell Center for the Space Sciences, Washington University in St. Louis, One Brookings Drive, St. Louis, MO 63130, USA}
\affiliation[d]{Dr. Karl Remeis-Observatory and Erlangen Centre for Astroparticle Physics, Universität Erlangen-Nürnberg, Sternwartstr. 7, D-96049 Bamberg, Germany}
\affiliation[e]{Deutsches Elektronen-Synchrotron (DESY), D-15738 Zeuthen, Germany}

\onbehalf{for the TELAMON collaboration} 

\emailAdd{jonas.hessdoerfer@uni-wuerzburg.de}

\abstract{We present recent results of the TELAMON program, which is using the Effelsberg 100-m telescope to monitor the radio spectra of active galactic nuclei (AGN) under scrutiny in astroparticle physics, namely TeV blazars and neutrino-associated AGN. 
Our sample includes all known Northern TeV-emitting blazars as well as blazars positionally coincident with IceCube neutrino alerts.
Polarization can give additional insight into the source properties, as the polarized emission is often found to vary on different timescales and amplitudes than the total intensity emission.
Here, we present an overview of the polarization properties of the TeV-emitting TELAMON sources at four frequencies in the $\SI{20}{mm}$ and $\SI{7}{mm}$ bands.
While at $\SI{7}{mm}$ roughly $82\,\%$ of all observed sources are found to be significantly polarized, for $\SI{20}{mm}$ the percentage is $\sim58\,\%$.
We find that most of the sources exhibit mean fractional polarizations of $<5\%$, matching the expectations of rather low polarization levels in these sources from previous studies at lower radio frequencies.
Nevertheless, we demonstrate examples of how the polarized emission can provide additional information over the total intensity.

}

\ConferenceLogo{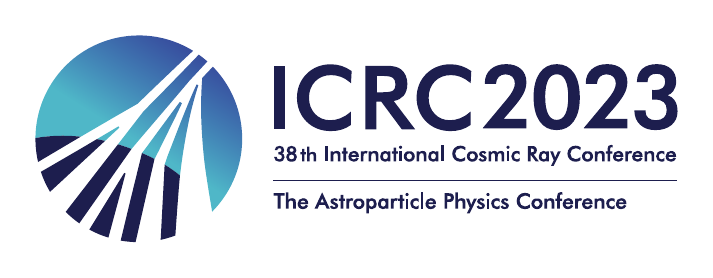}

\FullConference{%
38th International Cosmic Ray Conference (ICRC2023)\\
  26 July - 3 August, 2023\\
  Nagoya, Japan}

\begin{document}
\maketitle

\section{Introduction}
\noindent
Blazars are radio-loud active galactic nuclei (AGN) hosting relativistic jets pointed close to our line of sight. 
Their emission is highly beamed and Doppler-boosted which makes them variable broadband emitters from radio to $\gamma$-ray energies.
Their spectral energy distribution (SED) shows a double-humped structure. 
With decreasing luminosity, the peaks of blazar SEDs are shifted towards higher frequencies and the high-energy emission reaches the very-high-energy (VHE) regime at TeV $\gamma$-rays.
While the first component corresponds to synchrotron emission, there are two different explanations for the high-energy emission of blazars: inverse Compton scattering and hadronic emission models.
In the latter, the high-energy $\gamma$-ray emission is produced by interactions of relativistic protons in the jet with soft ambient seed photons \cite{mannheim1993proton} and bright neutrino emission is naturally expected.
The class of blazars is further divided into low-, intermediate-, and high-synchrotron peaked (LBL, IBL, HBL) sources according to their synchrotron peak frequency, where HBLs are canonically defined as sources whose synchrotron emission hump peaks above $10^{15}\,\mathrm{Hz}$ \cite{padovani1994connection}.
In the most extreme cases (extreme blazars or EHBLs), the peak of the synchrotron emission can reach even higher frequencies by up to two orders of magnitude \cite{biteau2020progress}.


\noindent
A better understanding of the extreme behavior is promised to arise from more comprehensive observations of extreme blazars and other TeV-blazars in the radio band, especially at high frequencies where one is probing the most compact jet emission regions.
TeV blazars seem to be very efficient in tapping bulk kinetic energy of their black hole jets (e.g., \cite{piner2018multi}), which makes them a key class of objects in the field of astroparticle physics as the potentially dominant source of ultrahigh-energy cosmic rays \cite{hillas1984origin}.
Similarly, TeV blazars are considered to contribute significantly to the highest-energy neutrinos detected (\cite{tavecchio2014structured}, \cite{padovani2015simplified}, \cite{giommi2020dissecting}).
Due to their high peak frequencies, HBL blazars are generally faint radio sources, which makes it hard to detect them with single-dish observations at high radio frequencies.
The TELAMON (Tev Effelsberg Long-term Agn MONitoring) program (\cite{kadler2021telamon}, Eppel et al. in prep) aims to characterize the variability of the radio spectra of AGN with VHE astroparticle emission.
More specifically, we target all 59 known TeV-detected AGN in the Northern Hemisphere (i.e., Decl. $>\SI{0}{\degree}$) as well as candidate neutrino-emitting AGN\footnote{A full source list is available on \url{https://telamon.astro.uni-wuerzburg.de/sources}.}.
We perform observations of these sources roughly every four weeks at multiple high radio frequencies up to $\SI{45}{GHz}$ and record total intensity as well as polarized intensity data.\\
The latter is of special interest, as previous studies (e.g., \cite{aller1996centimeter}, \cite{agudo2018polami}) found the variability of the polarized intensity to be faster and its fractional amplitude to be higher than for the total intensity.
With the addition of the polarized emission of the source, it is possible to super-resolve potential flares, allowing for an overall better modeling of the variability.\\
Here, we present first results of the polarization characteristics of the TELAMON sample.


\section{Observations and Data Analysis}
\noindent
For this work, we consider observations conducted within the TELAMON program in the time frame between September 2021 to June 2023, using the Effelsberg 100-m telescope that is operated by the Max-Planck-Institut für Radioastronomie (MPIfR).
Generally, TELAMON collects total intensity data in four different wavelength bands centered around $\SI{45}{mm}$, $\SI{20}{mm}$, $\SI{14}{mm}$ and $\SI{7}{mm}$ that are split into a total of twelve frequency bands of $\SI{2.5}{GHz}$ width each.
In order to maximize their detection rates, we observe sources in different wavelength bands depending on their flux densities.
However, this also means that we do not observe all sources at all frequencies, but rather observe the faint sources at low frequencies and the bright sources at the highest frequencies.
As all our sources appear as point sources to the telescope, we perform \enquote{cross-scans} over the source position in two perpendicular directions (azimuth and elevation) and record the antenna temperature.
We perform quality checks of these scans using a semi-automated flagging algorithm implemented by Eppel et al. (in prep), correct them for pointing offsets, atmospheric opacity and elevation-dependent gain effects, based on \cite{angelakis2019f}.
As calibrators, we use 3C\,286, NGC\,7027 and W3OH to convert the measured antenna temperatures to jansky, using the models of \cite{perley2017accurate}, \cite{zijlstra2008evolution} and our own model based on a free-free emission model and archival data, respectively for the three sources.\\
Here, we focus on the observations conducted using the \enquote{SpecPol}-backend of these receivers.
More specifically, we present the data taken in the bands centered around $\SI{14}{GHz}$, $\SI{17}{GHz}$, $\SI{36}{GHz}$ and $\SI{39}{GHz}$ of the $\SI{20}{mm}$ and $\SI{7}{mm}$ receivers.
We use frequency masks in order to avoid disturbances at the edges of the bands as well as radio frequency interference.
Thus, we integrate over an effective bandwidth of $\sim\SI{1.9}{GHz}$.
In contrast to the regular backends of the receivers, the SpecPol backend also collects polarization data.
The receivers are equipped with circularly polarized feeds that record the LCP and RCP components of the wave.
We follow the definition of the Stokes parameters in a circular base as defined in \cite{myserlis2018full}.
For this publication, we focus only on the linear polarization, i.e., Stokes $Q$ and $U$, of the sources, mainly due to three reasons: 1) Blazars typically are not expected to emit circularly polarized radiation, 2) Circularly polarized feeds generally favor the measurement of linear polarization, 3) Stokes V calibration needs a considerable amount of extra work.
Therefore, in the following, we denote the Stokes vector as $\mathbf{S} = (I, Q, U)$, setting $V=0$.
From these quantities, the polarization properties of the received signal can be further quantified by the means of the intensity of linear polarization $p_\mathrm{lin}$, the fractional polarization $m_l$ and the electric vector position angle (EVPA) $\chi$ via
\begin{align}
    p_\mathrm{lin} &=\sqrt{Q^2+U^2}\\
    m_l &= p_\mathrm{lin} / I \\ 
    \chi &= \frac{1}{2}\arctan \frac{U}{Q}
\end{align}
with $\SI{0}{\degree} \leq \chi \leq \SI{180}{\degree}$.
When measuring the Stokes parameters, one has to consider the fact that due to the azimuthal mounting of the Effelsberg telescope the source seems to rotate with respect to a default reference frame over time, leading to a rotation of the EVPA.
Additionally, the imperfect receiving system introduces spurious instrumental polarization into the measurement, causing intrinsically unpolarized sources to appear polarized.
One calibration scheme to account for both of these effects is the Müller matrix method \cite{turlo1985calibration,kraus2003intraday}.
The basic idea is that the transfer function between the true Stokes parameters $\mathbf{S}_\mathrm{true}$ and the observed ones $\mathbf{S}_\mathrm{obs}$ is given by the elements of the Müller matrix $\mathcal{M}$.
This matrix therefore describes the effects the instrument has on the measured signal.
Together with a rotation matrix $\mathcal{R}$ that rotates the observed Stokes parameters to a common reference frame based on the parallactic angle $q$ of the source at the time of measurement, the true Stokes parameters can be expressed as
\begin{equation}
    \begin{pmatrix}
        I\\ Q \\ U
    \end{pmatrix}_\mathrm{obs} = \mathcal{M}\cdot\mathcal{R}\cdot \mathbf{S}_\mathrm{true} =
    \begin{pmatrix}
        M_{11} & M_{12} & M_{13}\\
        M_{21} & M_{22} & M_{23}\\
        M_{31} & M_{32} & M_{33}
    \end{pmatrix} 
    \begin{pmatrix}
        1 & 0 & 0\\
        0 & \cos2q & \sin2q\\
        0 & -\sin2q & \cos2q
    \end{pmatrix}
    \begin{pmatrix}
        I\\ Q \\ U
    \end{pmatrix}_\mathrm{true}.
\end{equation}
While determining $\mathcal{R}$ is straightforward, determining $\mathcal{M}$ is generally not.
In order to do so, at least three independent measurements of sources with well-known polarization properties are required to solve this system of equations unambiguously.
For TELAMON, we use the aforementioned calibrators, using the model by \cite{perley2013integrated} for 3C\,286, while NGC\,7027 and W3OH, a planetary nebula and a star forming region, respectively, are assumed to not emit any polarized radiation.
Once the Müller matrix is known for the calibration sources, its inverse, $\mathcal{M}^{-1}$, can be applied to the observed Stokes parameters of the target sources to get rid of the instrumental effects.
In the aforementioned time range, we managed to recover polarization data in a total of 32 distinct epochs.
As the influence of the polarized flux density on the total flux density should be relatively small due to their relative weakness, we only correct $Q$ and $U$ while keeping $I_\mathrm{obs}$ as is.
From the Müller matrix elements, we can define the instrumental polarization
\begin{equation}
    p_\mathrm{inst} = \frac{\sqrt{M_{21}^2+M_{31}^2}}{\left|M_{11}\right|}
\end{equation}
as a mean to express the overflow of total intensity to polarized emission.
This value has to be small in order for the data to make sense.
Indeed, we find average values of $0.50\,\%$, $1.62\,\%$, $0.74\,\%$ and $0.82\,\%$ for the $\SI{14}{GHz}$, $\SI{17}{GHz}$, $\SI{36}{GHz}$ and $\SI{39}{GHz}$ bands, respectively.

\section{Results}\noindent
In this section, we present some general results of the polarization behavior of the sources in the TELAMON sample as well as selected examples that showcase the ability of polarization information to support total intensity data and our ability to detect polarization even at high radio frequencies.

\subsection{Linear Polarization Distribution of TeV Emitting Blazars}
\begin{figure}[htp]
    \centering
    \includegraphics[width=0.6\linewidth]{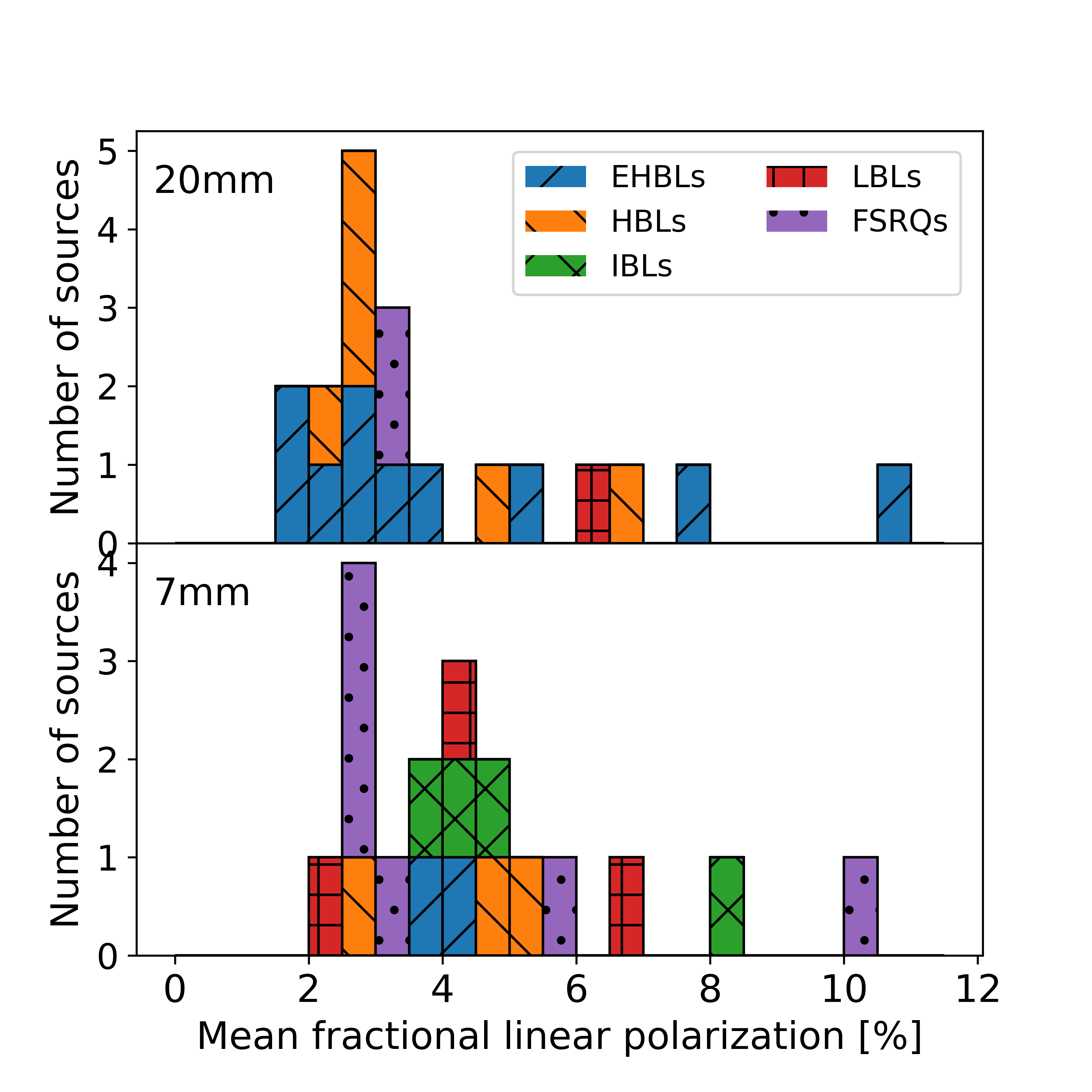}
    \caption{Distribution of the mean fractional linear polarization of all significantly polarized TeV emitting sources in the TELAMON sample.}
    \label{fig:lin_pol_distribution}
\end{figure}

\noindent
For this part of the results, we consider a source to be significantly polarized if it is polarized on a $2\sigma$ level, i.e., if it fulfills the condition $p_\mathrm{lin} > 2\sigma_{p\mathrm{lin}}$.
Additionally, we only consider the source detections by the SpecPol backend.
With this criterion, we find that $\sim58\,\%$ (at $\SI{20}{mm}$) and $\sim82\,\%$ (at $\SI{7}{mm}$) of the TeV-emitting sources in the sample are polarized.
For this statistic, we first averaged over the two individual frequencies of the respective receiver for all significantly polarized scans, and then averaged over all epochs with significant detections.
Here it is importatnt to note again that sources are not necessarily observed at the same frequencies.
In Fig.~\ref{fig:lin_pol_distribution}, we show the distribution of mean linear fractional polarization throughout our TeV sample for the different source types according to their source classification in Eppel et al. (in prep.).
From the figure, it is evident that at $\SI{20}{mm}$ most sources exhibit a mean fractional linear polarization of $<4\,\%$, while at $\SI{7}{mm}$ most sources are polarized at a level of $< 5\,\%$.
This generally matches well with the expectation of rather low polarizations in these types of sources (e.g., \cite{tabara1980catalogue}).
The by far largest group of sources that were not detected to be significantly polarized is made up of EHBLs, where only $\sim50\,\%$ of the sources were found to be polarized at $\SI{20}{mm}$. 
This makes sense, as these sources are the faintest sources in the sample and thus, on average, also exhibit the lowest polarized flux densities.
In fact, we found that sources with $p_\mathrm{lin} > \SI{0.0025}{Jy}$ at $\SI{20}{mm}$ were usually found to be significantly polarized, while sources with polarized flux densities below this level were usually not significantly detected.\\
Although most sources in the sample are only weakly polarized with some level of variability, there are a few sources that show continuous high polarization percentages.
At $\SI{20}{mm}$, the HBL MAGIC\,J2001+435 (J2001+4352) has a mean fractional polarization of $(6.8\pm 0.1)\,\%$ across eight observations, see Fig.~\ref{fig:sources} (middle).
The gravitationally lensed FSRQ S3\,0218+35 (J0221+3556) \cite{corbett1996radio} shows a relatively stable polarization percentage of $(10.0\pm 0.5)\,\%$ at $\SI{7}{mm}$, as is shown in Fig.~\ref{fig:sources} (right).
This is in agreement with the $\sim10\,\%$ polarization found by \cite{corbett1996radio} using $\SI{15}{GHz}$ VLA observations, suggesting that the source might be highly polarized throughout a wide range of radio frequencies.

\subsection{Selected examples}
\begin{figure}
    \centering
    \includegraphics[width=0.32\linewidth]{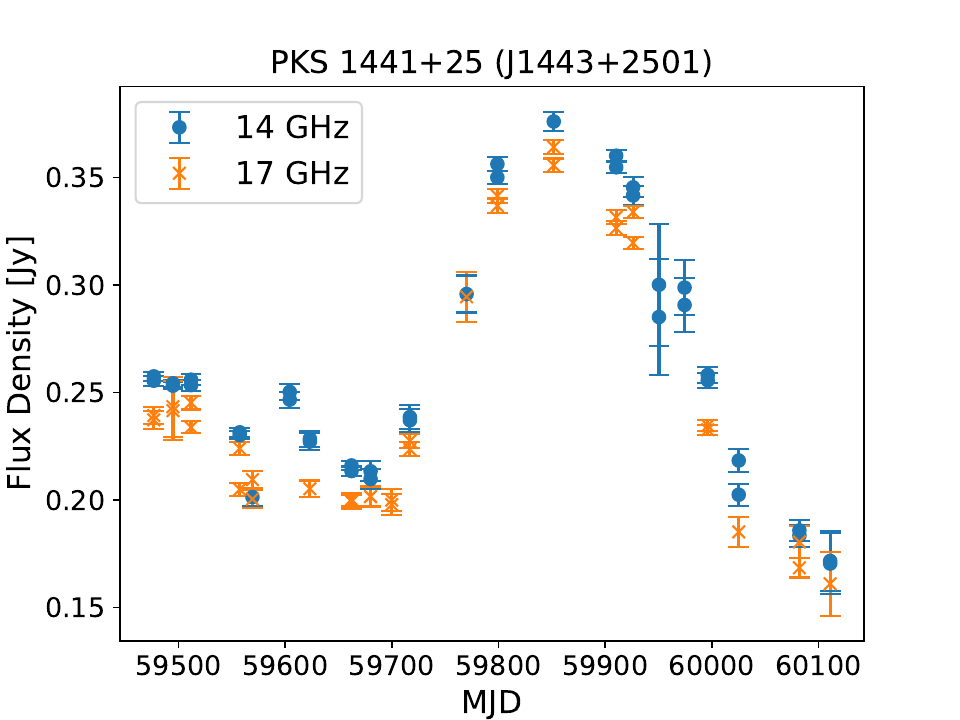}
    \includegraphics[width=0.32\linewidth]{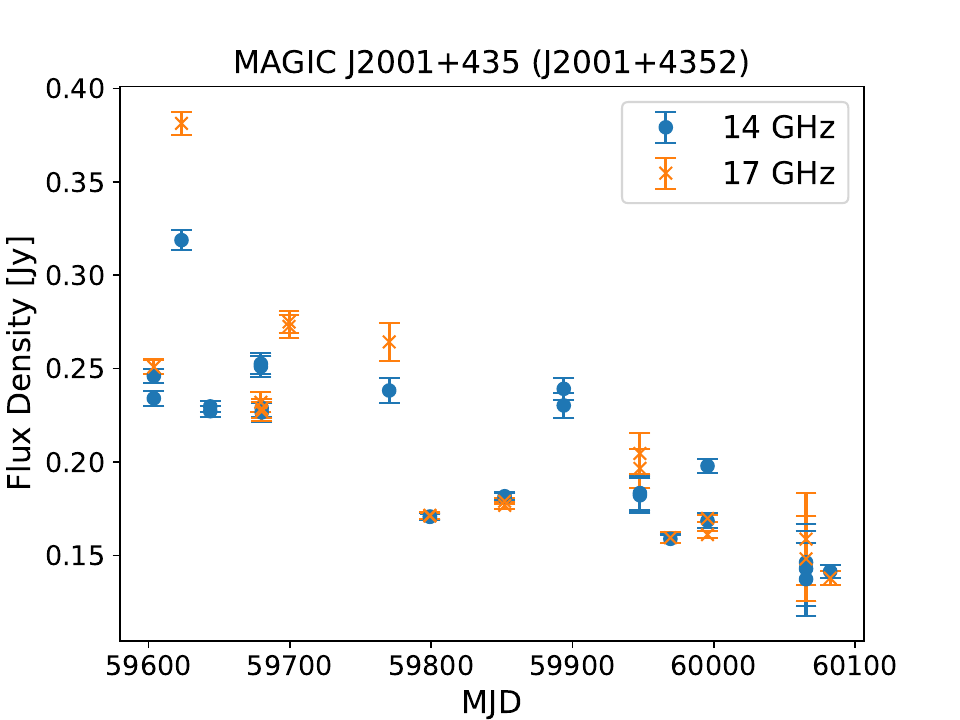}
    \includegraphics[width=0.32\linewidth]{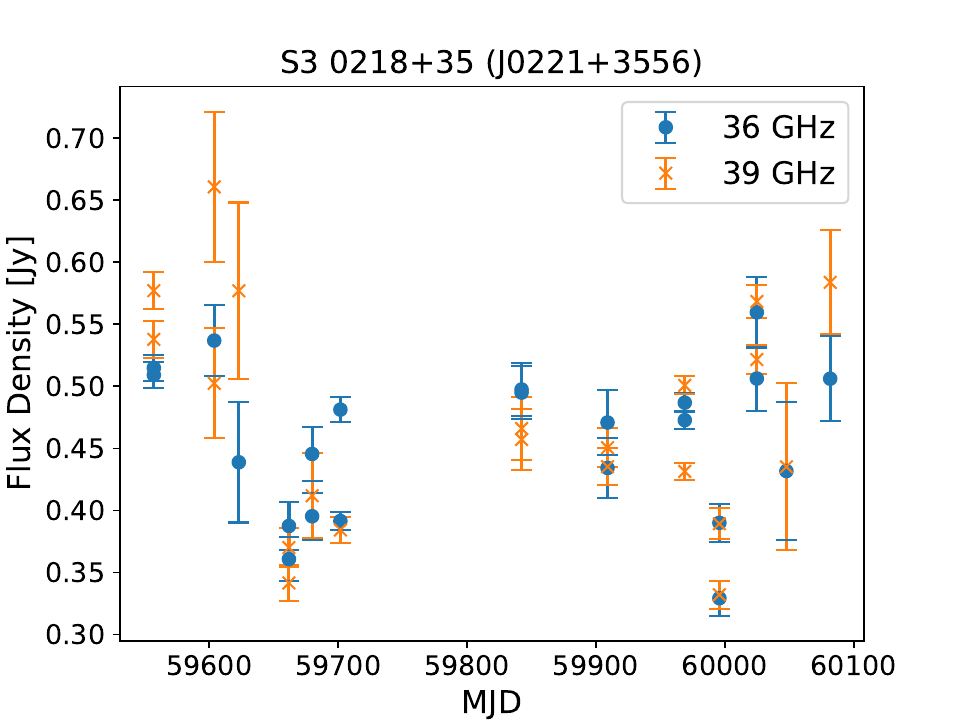}

    \includegraphics[width=0.32\linewidth]{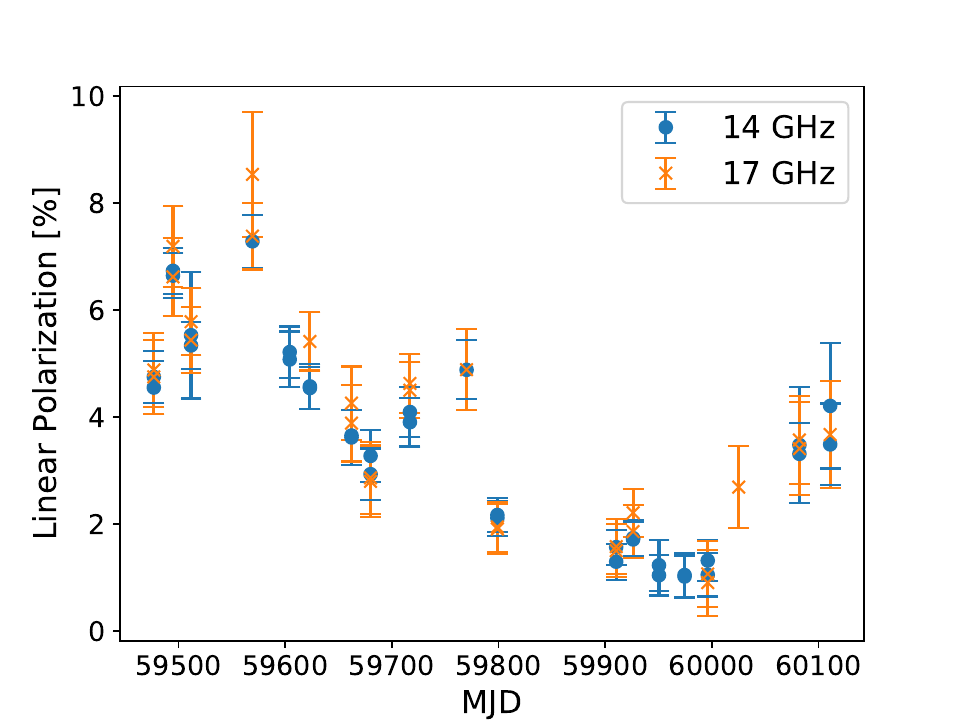}
    \includegraphics[width=0.32\linewidth]{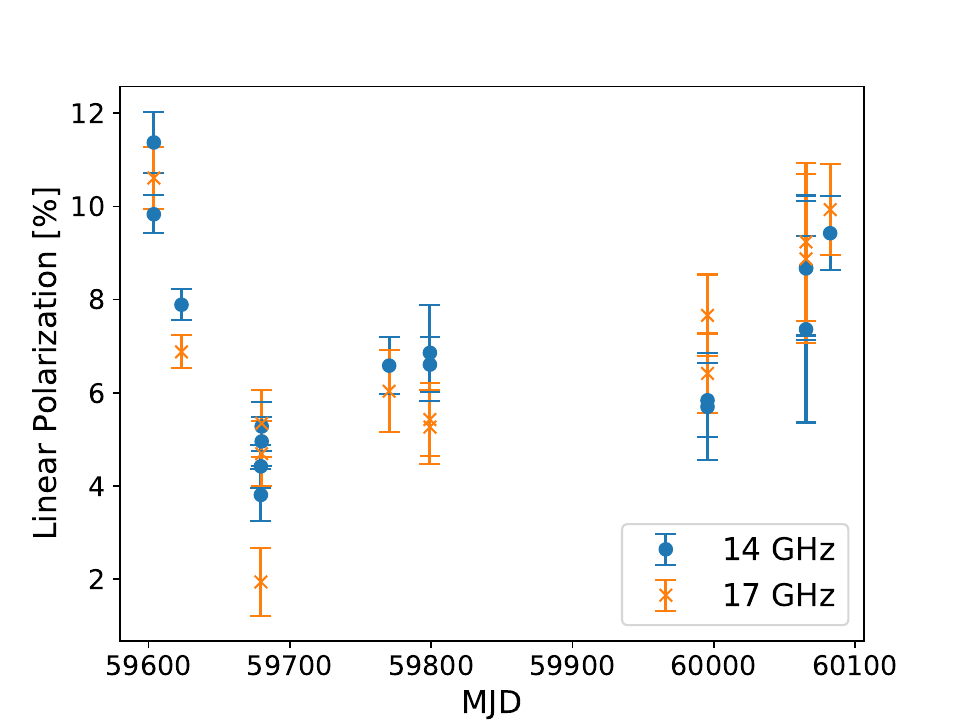}
    \includegraphics[width=0.32\linewidth]{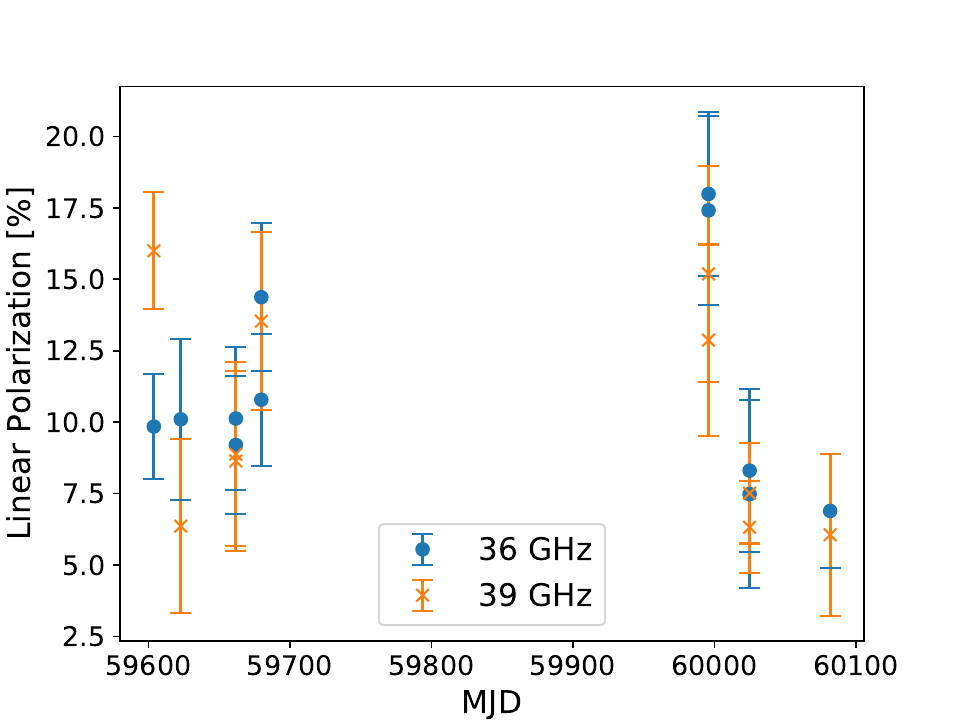}

    \includegraphics[width=0.32\linewidth]{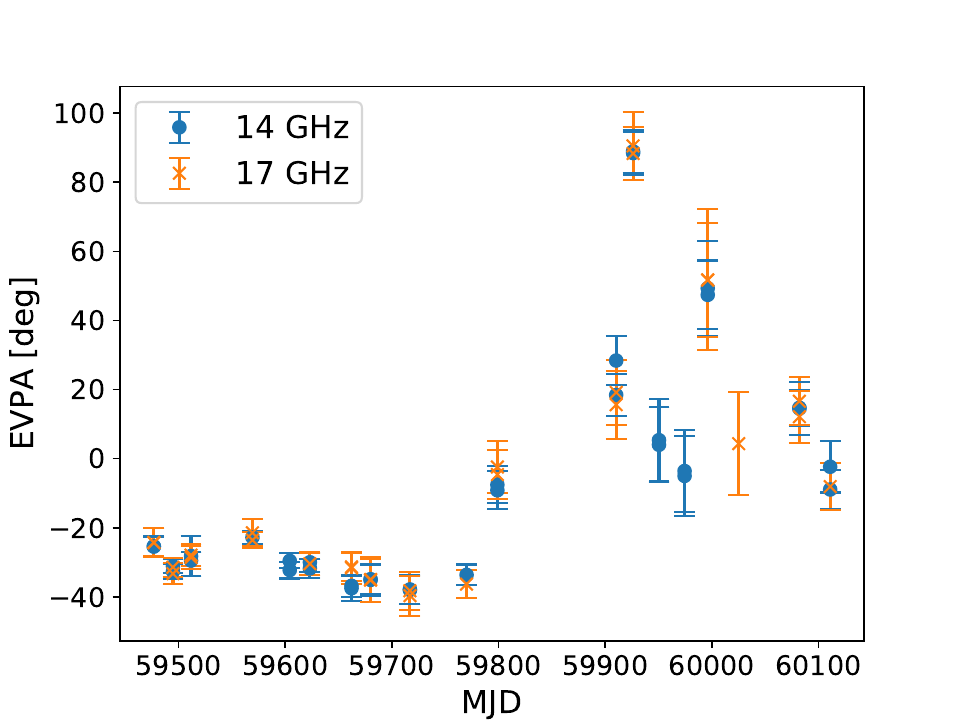}
    \includegraphics[width=0.32\linewidth]{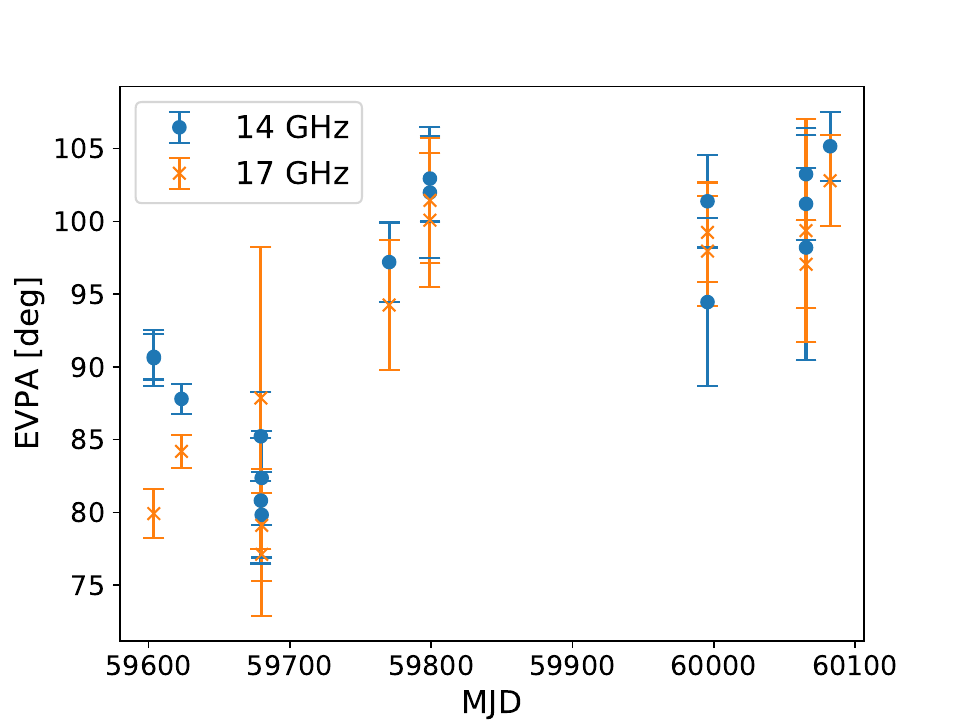}
    \includegraphics[width=0.32\linewidth]{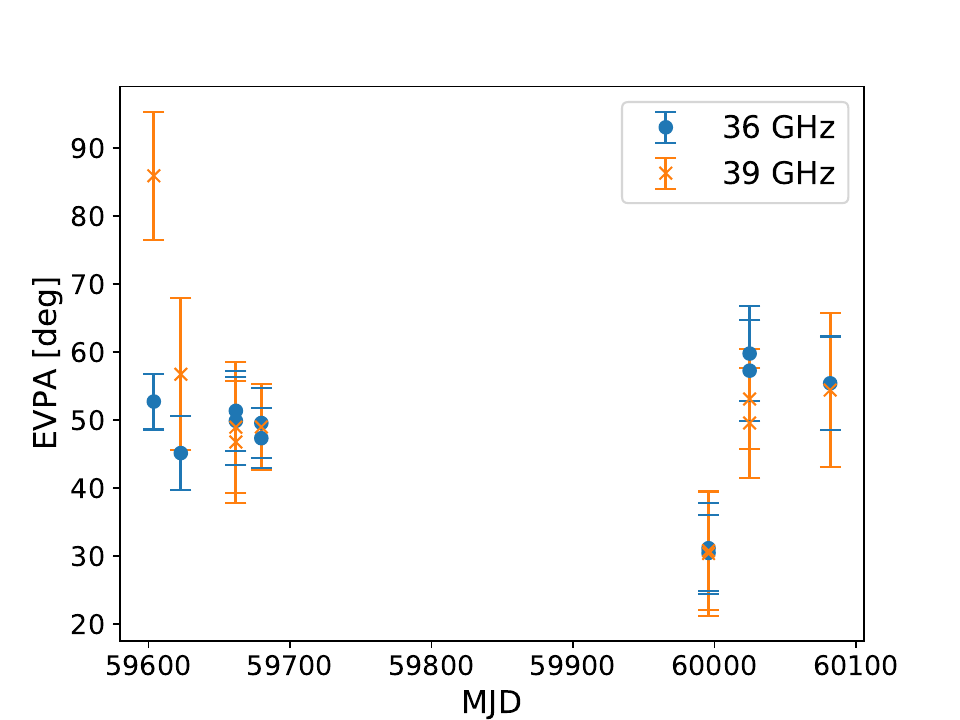}
    
    \caption{Total intensity (top row), linear polarization (center row) and EVPA (bottom row) evolution of the sources PKS\,1441+25 (J1443+2501), MAGIC\,J2001+435 (J2001+4352) and S3\,0218+35 (J0221+3556) (from left to right).
    All data points in this plot are significantly polarized.}
    \label{fig:sources}
\end{figure}

\noindent
Other than the overall sample, it is also worthwhile to have a look at specific sources to demonstrate the advantage of recording the polarization signals on top of the total intensity.
This is illustrated in Fig.~\ref{fig:sources} (left), where the FSRQ PKS 1441+25 (J1443+2501) is displayed.
The source showed a very prominent flare at $\SI{20}{mm}$ between MJD 59750 to 59950 that was accompanied by a drop in linear polarization percentage from a roughly constant $\sim4\%$ or higher to less than $2\%$.
After the flare ended, the polarization seems to be rising again, although not reaching the level before the flare yet, albeit at a fainter total intensity state.
At the same time as the rapid drop in polarization, the EVPA also becomes variable while being relatively stable beforehand.
The middle column of Fig.~\ref{fig:sources} shows a different kind of behavior, as MAGIC\,J2001+435 does not show a large variability in either the EVPA or the fractional linear polarization, while its total intensity emission seems to be steadily declining.
Lastly, the right column of Fig.~\ref{fig:sources} illustrates the source S3\,0218+35 at $\SI{7}{mm}$ frequencies that does not show any major variability in its total intensity emission.
However, around MJD 59700, it plateaus at a relatively low flux density for $\sim 250$\,days, while its linear polarization is very high $(>10\,\%)$ during this time. 
After the total intensity starts varying and rising again, the polarization declines to a moderately low value.
Unfortunately the sampling of the polarization during its high state is not dense due to technical difficulties with the SpecPol backends.
Nevertheless, this example nicely demonstrates that we are able to detect polarization also for modestly bright sources at high radio frequencies.

\section{Outlook}\noindent
This work represents a first look at the linear polarization properties of the TeV-blazars in the TELAMON sample.
While we focused mainly on the TeV-emitting sources here, further investigations of the neutrino candidate blazars could help to gain more information about potential systematic differences in their behaviors.
For this, we will look into refining our polarization analysis, possibly using a method similar to \cite{myserlis2018full} that does not rely on the Müller matrix formalism, but rather on simulating the instrumental effects to achieve better results for faint sources.
With this, we will also search for signatures of circular polarization in our data.
As TELAMON observes at four frequency bands, we will extend our study to the other bands as well.
Currently, we already have some data in the $\SI{45}{mm}$ band, however, it still needs some more checking as well as a larger sample size.
In our large data archive, we will look for correlations between the total intensity and polarized emission as well as the EVPA.
Finally, we will calculate rotation measures from the recorded EVPAs in order to gain insight into the magnetic field and particle densities along the line of sight to the sources.

\acknowledgments\noindent
This work is based on observations with the 100-m telescope of the MPIfR (Max-Planck-Institut für Radioastronomie) at Effelsberg.
FE, MK, FR and JH acknowledge support from the Deutsche Forschungsgemeinschaft (DFG, grants 447572188, 434448349, 465409577).

\bibliographystyle{JHEP}
\bibliography{References}

\clearpage
\section*{Full Authors List: TELAMON Collaboration}
%
%
\scriptsize
\noindent
U. Bach$^1$, 
P. Benke$^{1,6}$,
D. Berge$^2$,
S. Buson$^3$,
L. Debbrecht$^1$,
D. Dorner$^{4,6}$,
P.G. Edwards$^5$,
J. Eich$^6$,
F. Eppel$^{6,1}$,
C.M. Fromm$^6$,
M. Giroletti$^7$,
A. Gokus$^8$,
S. Hämmerich$^9$,
O. Hervet$^{10}$,
J. Heßdörfer$^{6,1}$,
M. Kadler$^6$,
A. Kappes$^1$,
D. Kirchner$^6$,
S. Koyama$^{11}$,
A. Kraus$^1$,
T.P. Krichbaum$^1$,
E. Lindfors$^{12}$,
K. Mannheim$^6$,
R. Ojha$^{13}$,
G.F. Paraschos$^1$,
E. Pueschel$^2$,
F. Rösch$^{6,1}$,
E. Ros$^1$,
B. Schleicher$^{4,6}$,
W. Schulga$^6$,
J. Sinapius$2$,
J. Sitarek$^{14}$,
P. Weber$^6$,
J. Wilms$^9$,
M. Zacharias$^{15,16}$,
and
J.A. Zensus$^1$

%

\noindent
$^1$Max-Planck-Institut für Radioastronomie, Auf dem Hügel 69, 53121, Bonn, Germany\\
$^2$Deutsches Elektronen-Synchrotron (DESY), 15738 Zeuthen, Germany\\
$^3$Julius-Maximilians-Universität Würzburg, Fakultät für Physik und Astronomie, Emil-Fischer-Str. 31, D-97074 Würzburg, Germany\\
$^4$Department of Physics, ETH Zurich, Otto Stern Weg 5, CH-8093 Zurich, Switzerland\\
$^5$CSIRO Space and Astronomy, PO Box 76, Epping, NSW, 1710, Australia\\
$^6$Julius-Maximilians-Universität Würzburg, Fakultät für Physik und Astronomie, Institut für Theoretische Physik und Astrophysik, Lehrstuhl für Astronomie, Emil-Fischer-Str. 31, D-97074 Würzburg, Germany\\
$^7$INAF-Istituto di Radioastronomia, Bologna, Via Gobetti 101, 40129, Bologna, Italy\\
$^8$Department of Physics \& McDonnell Center for the Space Sciences, Washington University in St. Louis, One Brookings Drive, St. Louis, MO 63130, USA\\
$^9$Dr. Karl Remeis-Observatory and Erlangen Centre for Astroparticle Physics, Universität Erlangen-Nürnberg, Sternwartstr. 7, 96049 Bamberg, Germany\\
$^{10}$Santa Cruz Institute for Particle Physics and Department of Physics, UCSC, Santa Cruz, CA 95064, USA\\
$^{11}$Niigata University, 8050 Ikarashi-nino-cho, Nishi-ku, Niigata 950-2181, Japan\\
$^{12}$Finnish Centre for Astronomy with ESO, University of Turku, FI-20014 University of Turku, Finland\\
$^{13}$NASA HQ, Washington, DC 20546, USA\\
$^{14}$Department of Astrophysics, Faculty of Physics and Applied Informatics, University of Łódź , ul. Pomorska 149/153, 90-236 Łódź, Poland\\
$^{15}$Landessternwarte, Universität Heidelberg, Königstuhl 12, 69117 Heidelberg, Germany\\
$^{16}$Centre for Space Research, North-West University, Potchefstroom 2520, South Africa


\end{document}